\documentclass{aastex631}

\shorttitle{Milky Way thin and thick disk kinematics with GAIA EDR3 and RAVE DR5}
\shortauthors{Vieira et al.}
\graphicspath{{./}{figures/}}
\begin{document}

\title{Milky Way thin and thick disks kinematics with GAIA EDR3 and RAVE DR5}

%\correspondingauthor{August Muench}
%\email{greg.schwarz@aas.org, gus.muench@aas.org}

\author[0000-0001-5598-8720]{Katherine Vieira}
\affiliation{Instituto de Astronom\'ia y Ciencias Planetarias, Universidad de Atacama, Copayapu 485, Copiap\'o 1531772, Chile}

\author[0000-0002-0155-9434]{Giovanni Carraro}
\affiliation{Dipartimento di Fisica e Astronomia, Universitá di Padova, Vicolo Osservatorio 3, I-35122 Padova, Italy}

\author[0000-0003-0250-6905]{Vladimir Korchagin}
\affiliation{Southern Federal University, Stachki 194, Rostov-on-Don 344090, Russia}

\author[0000-0002-5461-5778]{Artem Lutsenko}
\affiliation{Dipartimento di Fisica e Astronomia, Universitá di Padova, Vicolo Osservatorio 3, I-35122 Padova, Italy}

\author[0000-0002-5461-5778]{Terrence M. Girard}
\affiliation{Department of Physics, Southern Connecticut State University, 501 Crescent Street, New Haven, CT 06515, USA}

\author[0000-0002-5461-5778]{William van Altena}
\affiliation{Department of Astronomy, Yale University, 52 Hillhouse Avenue, New Haven, CT 06511, USA}

\begin{abstract}
We present a detailed analysis of kinematics of the Milky Way disk in solar neighborhood using GAIA DR3 catalog.  To determine the local kinematics of the stellar disks of the Milky Way galaxy we use a complete sample of 278,228 red giant branch (RGB) stars distributed in a cylinder, centered at the Sun with a 1 kpc radius and half-height of 0.5 kpc.  We determine separately the kinematical properties of RGB stars for each Galactic hemisphere in search for possible asymmetries. The kinematical properties of the RGB stars reveal the existence of two kinematically distinct components: the thin disk with mean velocities  ${V_R}, {V_{\phi}}, {V_Z}$  of about  -1, -239, 0 km s$^{-1}$ correspondingly and velocity dispersions $\sigma_R, \sigma_{\phi}, \sigma_Z$ of 31, 20 and 11 km s$^{-1}$, and the Thick disk with mean velocities components of about +1, -225, 0 km s$^{-1}$, and velocity dispersions of 49, 35, and 22 km s$^{-1}$.  We find that up to 500 pc height above/below the galactic plane, Thick disk stars comprise about half the stars of the disk. Such high amount of RGB stars with Thick disk kinematics points at the secular evolution scenario origin for the Thick disk of the Milky Way galaxy.
\end{abstract}


\section{Introduction} \label{sec:intro}

Our Galaxy consists of a few observationally distinct components, which differ for their chemical abundance and their kinematic properties: the thin disk, the Thick disk\footnote{To facilitate the reading of thin and Thick labels of disk samples, we will write the Thick one starting in capital letter.}, the bulge, and the halo. Thick disks are observed in most of the disk galaxies, besides the Milky Way. The Milky Way Thick disk bears evidence about the early history of the Galaxy, and is regarded as a significant component for understanding the process of Galaxy formation.

The Milky Way Thick disk was discovered by \citet{gilmore1983new}. Significant effort has been made afterwards to understand the origin and the physical properties of the Milky Way thick disk. However, after almost four decades, there is no consensus yet regarding the origin of the Thick disk of the Milky Way. To explain the origin of the Thick disk, a few scenarios have been suggested. A natural assumption of the Thick disk origin is the dynamical heating of a pre-existing thin disk by some mechanism. The heating of the thin disk can be produced by the scattering of thin disk stars by giant molecular clouds \citep{spitzer1951possible}, by spirals or barred structures \citep{sellwood1984spiral}, or by minor mergers of small companion galaxies \citep{quinn1993heating}. Radial migration of stars can also work as a possible mechanism of the Thick disk formation \citep{Roskar2008riding, schonrich2009chemical}.

Another group of scenarios involves a major accretion and a merger of a large satellite galaxy. \citet{abadi2003simulations} suggested first that the Thick disk was formed by the stars originated from the disrupted satellite galaxy. Somewhat intermediate to the pictures of the {\it internal} and {\it external} origin of the Thick disk is the scenario of the dual origin of the Galactic Thick disk, where the accretion of a significant merger triggers a centrally concentrated burst of star formation that marks the end of the formation of the rotationally-supported in-situ Thick disk that began forming prior to the merger. \citep{Bovy12} advocate for a continuous vertical disc structure connected by different individual stellar populations that smoothly go from Thin to Thick with increasing age. The integrated vertical space density of all these mono-abundance population reveals itself as identifying the Thin and Thick disk \citet{RixBovy}. On the other hand, \citet{Bird13}  confirmed this finding by simulating the vertical mass density profiles of individual age cohorts that are progressively steeper for younger populations. The superposition of age cohorts in the solar annulus results indeed in a double-exponential profile compatible with that observed in the Milky Way star counts. In other words,  the “upside-down” evolution that \citet{Bird12,Bird13} expose in their simulations indicates that the Thick disk arises from continuous trends between stellar age and metallicity. Therefore, the whole disk structure as it appears now is the result of a continuous evolution of the pristine disk stellar population, and does not originate from a discrete merger event, secular heating, or stellar radial migration after formation. 

Recent progress in the Thick disk studies is related to the high-resolution zoom-in cosmological simulations. Such simulations demonstrate that the cohorts of older stars have larger scale heights and shorter scale lengths, representing thus the Thick disk stellar population, while younger stars form the thin disk population of galaxies. In this picture the Thick disk is not a distinct component, but is rather a part of a double component system with a gradually-varying-with-height mixture of young and of old stars \citep{buck2020nihao}. \citet{park2021exploring} using high-resolution cosmological simulations GALACTICA and NEWHORIZON focused on the question whether the spatially defined thin and Thick disks are formed by different mechanisms. The authors traced the birthplaces of the stellar particles in the thin and Thick disks and found that most of the Thick disk stars in simulated galaxies were formed close to the mid-plane of the galaxies. This suggests that the two disks are not distinct in terms of the formation process but are rather the signature of a complex evolution of galaxies.

The only way to clarify which of the proposed formation scenarios was in place is a detailed study of the properties of the Milky Way Thick disk. To select Thick disk stars a few criteria have been suggested. \citet{fuhrmann2000deep} selected Thick disk stars using the relative abundance of alpha-elements of the stars taking into account the considerable difference in the abundances of such elements in the Milky Way thin and Thick disks. Attempts were made to distinguish each population based on the fact that the Thick disk is an older subsystem of the Galaxy while the thin disk stars are relatively young \citep{fuhrmann2000deep}. However, due to large errors in the determination of the age of individual stars this criterion seems rather unreliable.  The majority of studies of the properties of the Thick disk are based on the selection of Thick disk stars located at 1-2 kpc above or below the Galactic mid-plane where, presumably, most of the stellar population is represented by the Thick disk \citep{girard2006abundances, kordopatis2011spectroscopic}. In this study we use a different approach to select Thick disk stars.  The stars that belong to the Thick disk are located not only above or below the Galactic thin disk, but they are present as well in the solar neighborhood close to the Galactic mid-plane. Moreover, the concentration of Thick disk stars is expected to be highest near the mid-plane of the Galactic disk. Using kinematical data for the stars in solar neighborhood we can choose the stars that have relatively high velocities in the direction perpendicular to the galactic plane, so they will leave the solar neighborhood in a near future. The approach has a few obvious advantages. Kinematical data of the nearby stars are determined with a better accuracy compared to the kinematics of the distant stars. The volume density of the Thick disk stars is expected to be highest close to the mid-plane of the of Galaxy, decreasing exponentially perpendicular to the Galactic disk. Thus, selecting Thick disk stars in solar neighborhood in this way we obtain richer stellar sample with better determined kinematical properties.

As a consequence, this study is organized as follows: section 2 explains how the data samples were selected and their completeness assessed, section 3 is dedicated to the statistical calculation and analysis of the velocities' mean and dispersion, and the number density ratio of the Milky Way thin and Thick disks. Section 4 provides a brief discussion and section 5 summarizes our conclusions. An appendix is finally included with several additional explanations.

\section{Data selection and completeness} \label{sec:datasel}
We based our study on data taken from the Gaia Early Data Release 3 (EDR3) catalog \citep{2016A&A...595A...1G, 2021A&A...649A...1G} selecting a complete sample of good quality red giant branch (RGB) stars within the cylinder centered at Sun that has $\pm$ 0.5 kpc height and 1 kpc radius. A first selection from the Gaia catalog was done by imposing the following criteria: astrometric quality \texttt{ruwe}$<1.4$, parallax errors less than 10\%, apparent magnitudes $G<13.5$, and \texttt{parallax}$>0.89$ (4,863,317 sources). A further cut was done to select all sources within the above mentioned cylinder volume (4,274,340 sources), and having not null \texttt{bp\_rp} and \texttt{dr2\_radial\_velocity}, that is measured $B-R$ color and Gaia radial velocity (2,387,462 sources; $\sim 56$\% of cylinder sample). The value of \texttt{ruwe} is a renormalized unit-weight error which indicates how good is the astrometric solution of the star. Following the general recommendation from \citet{2021gdr3.reptE....V} we use the value 1.4 as a safe upper limit to select well-behaved data. Parallax error cut assures we can use the inverse of the parallax as a unbiased measure of the star distance, as explained by \citet{2021gdr3.reptE....V}. As described by the GAIA collaboration, GAIA EDR3 has in general twice-better proper motions and parallaxes compared to the GAIA DR2 catalog. Radial velocities in GAIA EDR3 catalog however, are just a copy of the radial velocities from GAIA DR2 catalog and are available only for a portion of stars with apparent magnitude $\sim 4<G<\sim 13$. 

\begin{figure}[ht!]
\plottwo{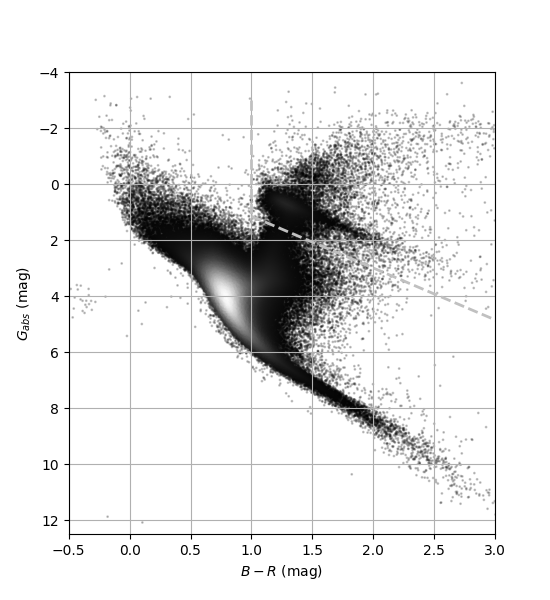}{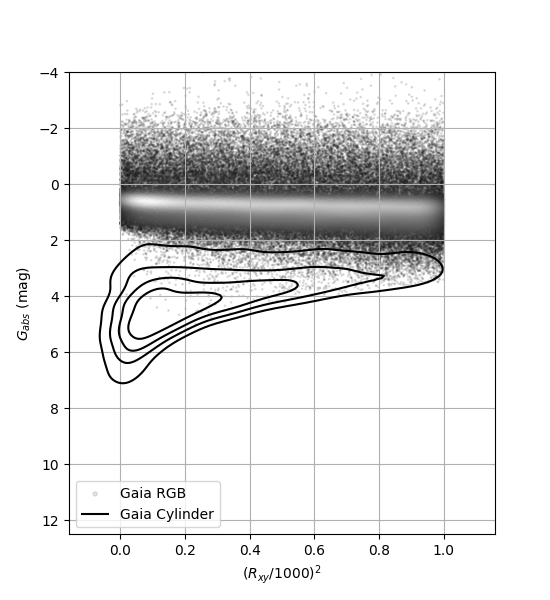}
\caption{Left panel: $G_{abs}$ vs. $B-R$ for all the Gaia data in the cylinder studied in this investigation, colour-coded by density of points (the whiter the denser, every 20 stars plotted). Our Gaia RGB sample was selected from the area delimited by the grey dashed line. Right panel:  $G_{abs}$ vs. $(R_{xy}/1000)^2$ for our Gaia RGB sample, colour-coded by density as in left panel (every 2 stars plotted). Grey contours are the Kernel Density Estimation of all the Gaia data in the cylinder (iso-density curves from 20\% to 80\% levels).} \label{fig:cmd}
\end{figure}

The plot of absolute magnitude $G_{abs}$ vs $B-R$ for the latter sample shows the presence of the red clump (RC) stars, as an over-density at $G_{abs}\sim 0.5$ on its blue end and tilted towards fainter $G$ magnitudes and redder $B-R$ colours due to the effect of interstellar extinction. The selection $B-R>1$ and $G_{abs}-1.85(B-R)<-0.7$ allowed us to choose all the RGB stars from the RC upwards (see Figure \ref{fig:cmd}, left panel), yielding a sample of 278,228 stars named Gaia RGB in Table \ref{tab_basic_samples}. This sample represents 94\% of all stars listed in Gaia EDR3 located within the cylinder and having the same color and magnitude selection of our sample (296,879 stars), meaning we are missing only 6\% of what Gaia observed in that volume for that kind of stars, due to lacking radial velocity information. 

A plot of $G_{abs}$ vs. distance (Figure \ref{fig:cmd}, right panel) shows the RC stars being well sampled up to 1.15 kpc in distance by our RGB sample, the only signature of incompleteness comes from the bright limit in Gaia, visible as the curved brighter end of the plotted data, meaning our sample is missing some nearby ($\lesssim 200$ pc) RC stars only. By selecting the red giant population, which is presumably older and less affected by kinematical features associated with their birth place, our results will be more indicative of the overall disk dynamics. 

We use Topcat \citep{2005ASPC..347...29T} functions \texttt{astromXYZ, astromUVW} and \texttt{icrsToGal} to compute rectangular coordinates and velocities oriented such that X(U) points towards the Galactic center, Y(V) points towards the galactic rotation, and Z(W) points towards the North Galactic Pole. Finally, the Galacto-centric  velocities in the cylindrical coordinate system $(V_R,V_\phi,V_Z)$ were computed assuming the solar motion with respect to the Local Standard of Rest (LSR) of $(U,V,W)_\odot=(11.1,12.2,7.3)$ km s$^{-1}$ \citep{2010MNRAS.403.1829S} and a galactic rotation for the LSR of -244.5 km s$^{-1}$ \citep{2020A&A...644A..83F} with $V_R$-axis pointing away from the Galactic center, $V_\phi$-axis oriented against the galactic rotation at the LSR and $V_Z$-action directed towards the North Galactic Pole. In total, 133,218 stars were selected as RGB stars in the northern part of the cylinder $(b>0)$ and 145,010 in the southern one ($b\leq 0)$ with the total number of stars being 278,228. Selected samples were chosen according to their $V_Z$ values, so that they are dominated by the thin and Thick disk populations respectively. We call the thin disk sample those stars having $V_z$ velocities of $|V_z|<15$ and heights $|Z|<200$ pc, and the Thick disk sample as those with $40<|V_z|<80$ (see Table \ref{tab_basic_samples}).

Therefore we did not split these GAIA$\times$RAVE samples into northern and southern parts. Metallicity value $[Fe/H]$ (labeled $Met_K$ in RAVE) was used to select the thin disk stars as those having the described above kinematical properties and $[Fe/H]>-0.4$, while the Thick disk stars have the described above kinematical properties and  $-1<[Fe/H]<-0.4$. Table \ref{tab_basic_samples} summarizes the number of stars in each Gaia/Gaia$\times$RAVE, North/South, thin/Thick disk sample. Details on the Gaia$\times$RAVE cross-match are discussed in Appendix \ref{app_xmatch}.

\begin{table}[ht]
    \centering
    \begin{tabular}{p{6cm}rrr}
    Sample name & North & South & Total \\ \hline\hline
    Gaia RGB & 133,218 & 145,010 & 278,228 \\
    Gaia RGB thin disk  & 54,197 & 55,628 & 109,825 \\
    Gaia RGB Thick disk & 6,776 & 7,022 & 13,798 \\ \hline
    Gaia$\times$RAVE RGB & \multicolumn{3}{r}{30,170} \\
    Gaia$\times$RAVE RGB thin disk  & \multicolumn{3}{r}{3,381} \\
    Gaia$\times$RAVE RGB Thick disk & \multicolumn{3}{r}{1,113} \\ \hline
    \end{tabular}
    \caption{Samples names and sizes used in this work as described in Section \ref{sec:datasel}.}
    \label{tab_basic_samples}
\end{table}

\subsection{Completeness}\label{sec:datacomp}
\begin{figure}[ht!]
\plottwo{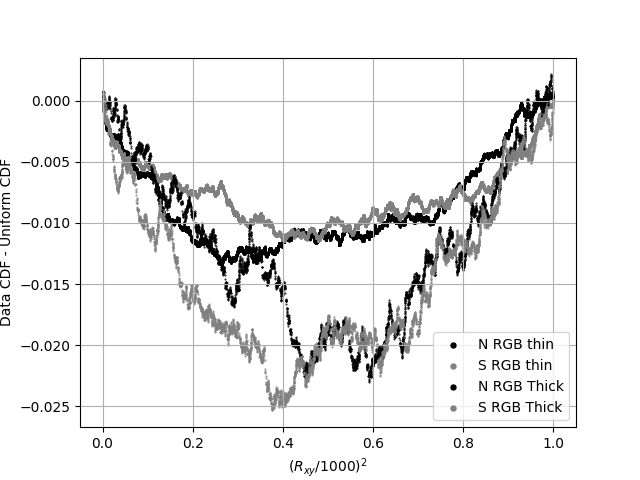}{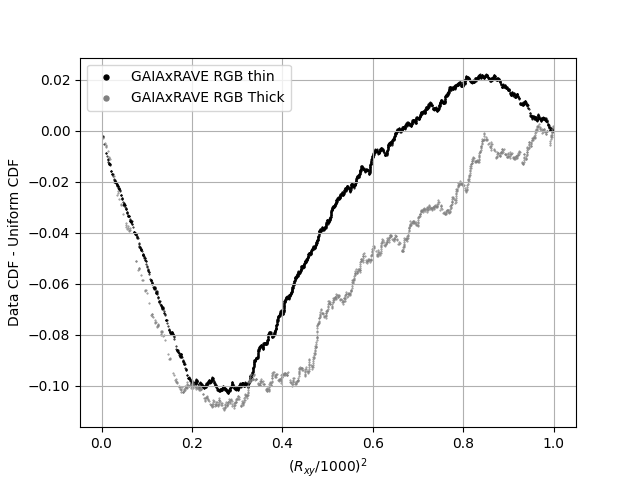}
\caption{Cumulative distribution function (CDF) of the normalized cylindrical radius squared minus the identity function, for the samples studied in this work.} \label{fig:cdf}
\end{figure}

Assuming that stellar density changes only vertically and is symmetrically distributed with respect to the Galactic plane, the empirical cumulative distribution function (CDF) of the normalized cylindrical radius squared $(R_{xy}/1000)^2=(X^2+Y^2)/1000^2$ should grow linearly, following very closely the identity function. In other words, in a complete sample, the number of stars that are uniformly distributed in a cylinder should grow proportionally to the square of the radius of the cylinder. The CDF is obtained by sorting the $(R_{xy}/1000)^2$ data and computing the percentile of each data, versus the data itself. A complete sampling is achieved when the CDF is very close to the identity function, i.e. the number of stars grows linearly with $R^2_{xy}$. Any deviation from this law means that the sample is not homogeneous or incomplete. Figure \ref{fig:cdf} plots the CDF minus the identity function to check how much the CDF deviates from the latter. For the samples studied in this work we find that the Gaia ones are essentially complete with incompleteness being about 1\%. Their CDF being below the identity means these Gaia stars should occupy higher percentiles, i.e. some stars are missing at smaller distances, and this is consistent with Figure \ref{fig:cmd} (right panel), that shows a slight incompleteness occurs at the fainter end of the RGB stars at smaller cylindrical radius distances.

The Gaia+RAVE samples deviate more from the identity function. For example the Gaia+RAVE thin disk overcomes the identity at $(R_{xy}/1000)^2\gtrsim 0.64$ (i.e. $R_{xy}\gtrsim 800$ pc), that is the data there radially accumulates faster than expected for a radially uniform distribution. Gaia+RAVE samples are about 5\% incomplete in radius sampling. Details on how incompleteness is computed are provided in Appendix \ref{app_complete}.

\begin{table}[ht]
    \centering
    \begin{tabular}{p{6cm}rr}
    Sample name & North & South  \\ \hline\hline
    Gaia RGB & 1.02 & 0.91  \\
    Gaia RGB thin disk  & 0.84 & 0.77  \\
    Gaia RGB Thick disk & 1.19 & 1.46  \\ \hline
    Gaia$\times$RAVE RGB & \multicolumn{2}{c}{4.87} \\
    Gaia$\times$RAVE RGB thin disk  & \multicolumn{2}{c}{3.36} \\
    Gaia$\times$RAVE RGB Thick disk & \multicolumn{2}{c}{5.84} \\ \hline
    \end{tabular}
    \caption{Samples incompleteness percentages, as described in Section \ref{sec:datacomp}.}
    \label{tab_complete}
\end{table}

\section{Kinematics of the thin and thick disks}
\subsection{Radial and azimuthal velocities}

\begin{figure}[ht!]
\plottwo{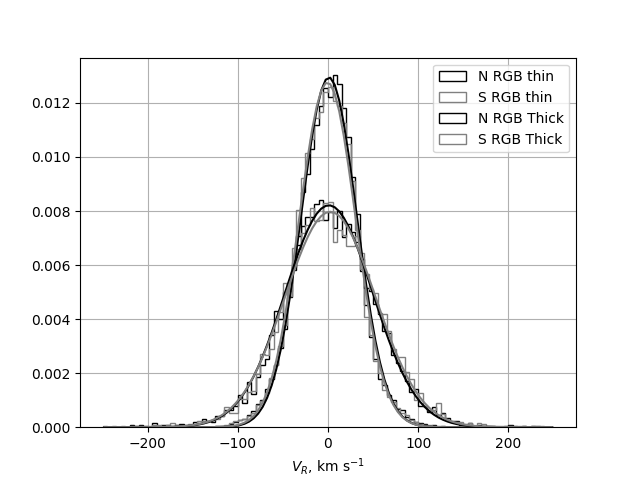}{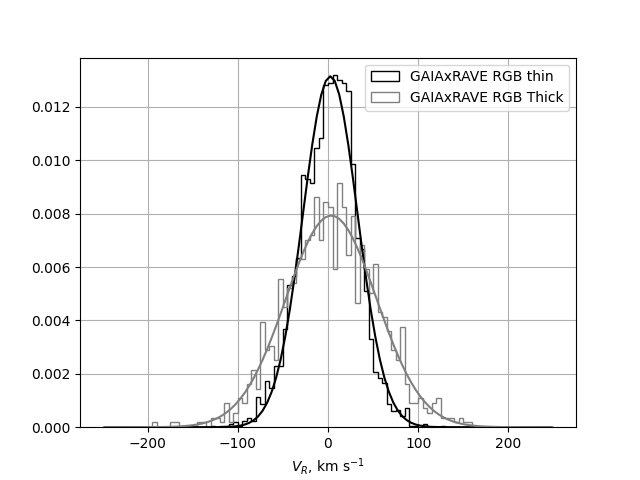}
\caption{Histograms and Gaussian fit of $V_R$ for the samples studied in this work. See also Tables \ref{tab_momvr} and \ref{tab_gfitvr} \label{fig:histovr}.}
\end{figure}

Figure \ref{fig:histovr} shows distributions of the radial velocity $V_R$ computed with 5 km s$^{-1}$ bins separately for the thin and Thick disks in the northern and southern parts of the cylinder. Left panel presents the radial velocity distributions measured for our complete samples of stars. Right panel presents the same distribution for the samples that also have the chemical abundance information. Similarity of both distributions proves that lack of chemical abundance information does not bias substantially the conclusions based on purely kinematical data. Mean values and standard deviations of the radial velocity distributions in all samples were computed using the standard algebraic expression (see Table \ref{tab_momvr}), and estimated also with help of a Gaussian fit from the Python {\tt\string scipy} package (see Table \ref{tab_gfitvr}). As one can see from the tables, mean radial velocities in all samples are of order of 1 - 2 km s$^{-1}$, in both thin and Thick samples, and radial velocity dispersions are $\sim$ 50 km $^{-1}$ for the Thick disk, and $\sim$ 30 km $^{-1}$ for the thin disk samples.

\begin{figure}[ht!]
\plottwo{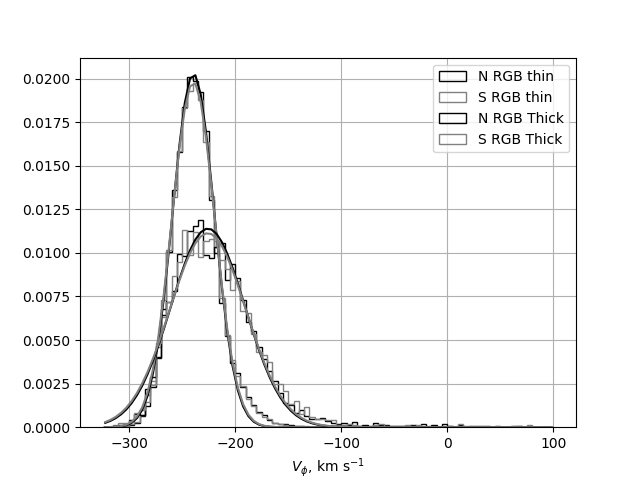}{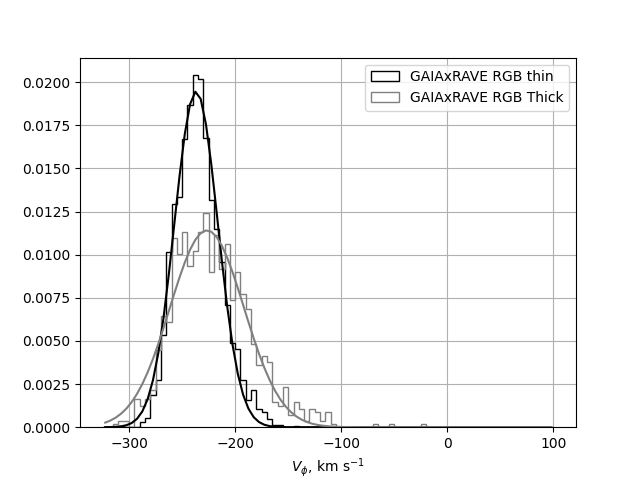}
\caption{Histograms and Gaussian fit of $V_\phi$ for the samples studied in this work. See also Tables \ref{tab_momvphi} and \ref{tab_gfitvphi}. \label{fig:histovphi}}
\end{figure}

Figure \ref{fig:histovphi} together with Tables \ref{tab_momvphi} and \ref{tab_gfitvphi}, present measurements of the azimuthal velocity mean and dispersion for selected samples of stars. As one can see, the $V_\phi$-velocity distribution not symmetric, showing the presence of asymmetric drift in the samples. Again, the kinematically selected samples have values comparable to those estimated with help the much smaller GAIA$\times$RAVE samples that include stellar abundance information. This proves that the lack of chemical information does not bias the results.

As one can see from the Tables \ref{tab_momvphi} and \ref{tab_gfitvphi}, mean velocities of the thin and of the Thick disks lag from the assumed velocity of the local standard of rest of -244.5 km s$^{-1}$. The thin disk sample is lagging behind LSR rotational velocity by 5 to 8 km s$^{-1}$. The Thick disk in the solar neighborhood is lagging behind rotational velocity of LSR by about 20 km s$^{-1}$. The observed lag of the rotational velocity in both samples is caused by two reasons. The lag of the rotational velocity in both subsystems is caused by asymmetric drift, the difference between the gravitational force and the centrifugal force caused by the nonzero components of the velocity dispersion of the systems \citep{binney2008galactic}. Another reason for the existence of asymmetric drift is that the thin and Thick disks are mixed, which will be proved in the next section by the analysis of kinematics of RGB stars in the direction perpendicular to the Galactic disk.

\begin{table}[ht]
    \centering
    \begin{tabular}{p{5cm}rcccc}
     & \multicolumn{2}{c}{Mean} & \multicolumn{2}{c}{StdDev}  \\ \hline
    Sample name & North & South & North & South \\ \hline\hline
    Gaia RGB thin disk & 0.01 $\pm$ 0.14 & -1.04 $\pm$ 0.14 & 32.65 $\pm$ 0.10 & 32.64 $\pm$ 0.10  \\
    Gaia RGB Thick disk & 1.67 $\pm$ 0.64 & 3.53 $\pm$ 0.63 & 52.54 $\pm$ 0.45 & 52.48 $\pm$ 0.44  \\ \hline
    Gaia$\times$RAVE RGB thin disk  & \multicolumn{2}{c}{0.48 $\pm$ 0.54} & 
    \multicolumn{2}{c}{31.29 $\pm$ 0.38} \\  
    Gaia$\times$RAVE RGB Thick disk & \multicolumn{2}{c}{4.00 $\pm$ 1.52} & \multicolumn{2}{c}{50.85 $\pm$ 1.08} \\ \hline
    \end{tabular}
    \caption{$V_R$ mean and standard deviation values for the samples studied in this work. Errors are computed as StdDev/$\sqrt{n-1}$ and StdDev/$\sqrt{2n}$ respectively.}
    \label{tab_momvr}
\end{table}

\begin{table}[ht]
    \centering
    \begin{tabular}{p{5cm}rcccc}
     & \multicolumn{2}{c}{$\mu$} & \multicolumn{2}{c}{$\sigma$} \\ \hline    
    Sample name & North & South & North & South \\ \hline\hline
    Gaia RGB thin disk & 1.05 $\pm$ 0.30 & -0.64 $\pm$ 0.27 & 30.81 $\pm$ 0.25 & 31.26 $\pm$ 0.22  \\
    Gaia RGB Thick disk & 0.98 $\pm$ 0.48 & 2.50 $\pm$ 0.57 & 48.58 $\pm$ 0.39 & 50.04 $\pm$ 0.47  \\ \hline
    Gaia$\times$RAVE RGB thin disk  & \multicolumn{2}{c}{2.51 $\pm$ 0.70} & 
    \multicolumn{2}{c}{30.33 $\pm$ 0.57} \\  
    Gaia$\times$RAVE RGB Thick disk & \multicolumn{2}{c}{3.56 $\pm$ 1.41} & \multicolumn{2}{c}{50.31 $\pm$ 1.15} \\ \hline
    \end{tabular}
    \caption{$V_R$ mean $\mu$ and dispersion $\sigma$ values from a Gaussian fit (Figure \ref{fig:histovr}) with their respective errors, for the samples studied in this work.}
    \label{tab_gfitvr}
\end{table}

\begin{table}[ht]
    \centering
    \begin{tabular}{p{5cm}rcccc}
     & \multicolumn{2}{c}{Mean} & \multicolumn{2}{c}{StdDev}  \\ \hline    
    Sample name & North & South & North & South \\ \hline\hline
    Gaia RGB thin disk & -236.80 $\pm$ 0.09 & -237.11 $\pm$ 0.09 & 22.06 $\pm$ 0.07 & 22.32 $\pm$ 0.07  \\
    Gaia RGB Thick disk & -218.07 $\pm$ 0.52 & -218.26 $\pm$ 0.50 & 43.07 $\pm$ 0.37 & 42.19 $\pm$ 0.36  \\ \hline
    Gaia$\times$RAVE RGB thin disk  & \multicolumn{2}{c}{-234.22 $\pm$ 0.37} & 
    \multicolumn{2}{c}{21.30 $\pm$ 0.26} \\  
    Gaia$\times$RAVE RGB Thick disk & \multicolumn{2}{c}{-221.13 $\pm$ 1.09} & \multicolumn{2}{c}{36.38 $\pm$ 0.77} \\ \hline
    \end{tabular}
    \caption{$V_\phi$ mean and standard deviation values for the samples studied in this work. Errors are computed as StdDev/$\sqrt{n-1}$ and StdDev/$\sqrt{2n}$ respectively.}
    \label{tab_momvphi}
\end{table}

\begin{table}[ht]
    \centering
    \begin{tabular}{p{5cm}rcccc}
     & \multicolumn{2}{c}{$\mu$} & \multicolumn{2}{c}{$\sigma$} \\ \hline    
    Sample name & North & South & North & South \\ \hline\hline
    Gaia RGB thin disk & -239.31 $\pm$ 0.23 & -239.40 $\pm$ 0.19 & 19.67 $\pm$ 0.19 & 20.15 $\pm$ 0.15  \\
    Gaia RGB Thick disk & -225.67 $\pm$ 0.70 & -225.80 $\pm$ 0.79 & 35.01 $\pm$ 0.57 & 35.84 $\pm$ 0.65  \\ \hline
    Gaia$\times$RAVE RGB thin disk  & \multicolumn{2}{c}{-236.82 $\pm$ 0.54} & 
    \multicolumn{2}{c}{20.50 $\pm$ 0.44} \\  
    Gaia$\times$RAVE RGB Thick disk & \multicolumn{2}{c}{-226.61 $\pm$ 1.31} & \multicolumn{2}{c}{34.97 $\pm$ 1.07} \\ \hline
    \end{tabular}
    \caption{$V_\phi$ mean $\mu$ and dispersion $\sigma$ values from a Gaussian fit (Figure \ref{fig:histovphi}) with their respective errors, for the samples studied in this work.}
    \label{tab_gfitvphi}
\end{table}

\subsection{Vertical velocities, density ratio and North-South symmetry}
\begin{figure}[ht!]
\plottwo{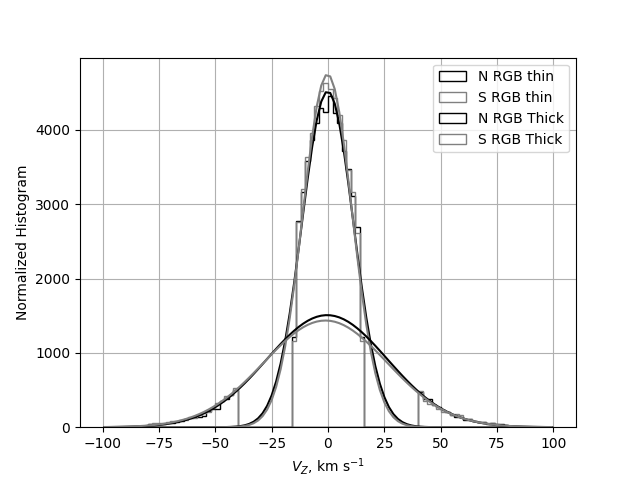}{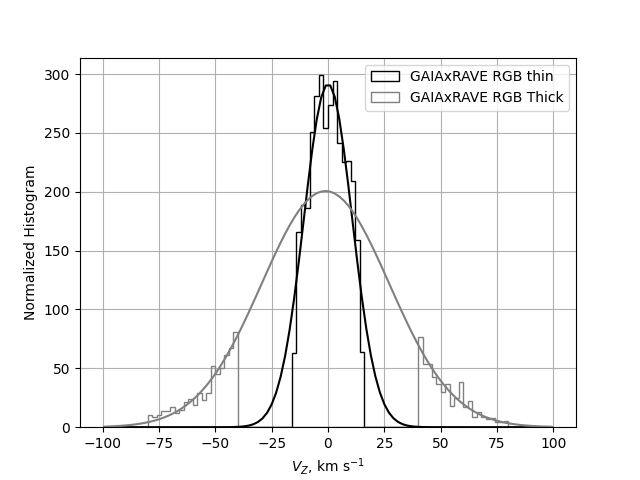}
\caption{Histograms and Gaussian fit of $V_Z$ for the samples studied in this work.
See also Table \ref{tab_gfitvz1}. \label{fig:histovz}}
\end{figure}

The distribution of vertical velocities $V_Z$ for the Gaia RGB sample cannot be fitted by a single Gaussian function. Our selection of the thin and Thick disk samples was respectively based on the $|V_z|<15$ km s$^{-1}$ {\it core} and the  $40<|V_z|<80$ km s$^{-1}$ {\it wings} of the $V_Z$ distribution. A zero-centered Gaussian fit corresponding to each of these portions in the non-normalized histogram is shown in Figure \ref{fig:histovz}. The obtained amplitude, mean and dispersion $(A,\mu,\sigma)$ values are shown in Table \ref{tab_gfitvz1}. These plots demonstrate that the thin disk sample can be heavily contaminated by Thick disk stars, although it must be kept in mind that the thin disk samples have $|Z|<200$ pc while the Thick disk samples go up to $|Z|<500$ pc. 

\begin{table}[ht]
    \centering
    \begin{tabular}{p{4.5cm}rcccccc}
    & \multicolumn{2}{c}{$\mu$} & \multicolumn{2}{c}{$\sigma$} & \multicolumn{2}{c}{$A$}  \\ \hline    
    Sample name & North & South & North & South & North & South \\ \hline\hline
    Gaia RGB thin disk & -0.14 $\pm$ 0.47 & -0.21 $\pm$ 0.41 
    & 11.44 $\pm$ 0.65 & 10.94 $\pm$ 0.55 & 129609 $\pm$ 5584 & 130213 $\pm$ 5008 \\
    Gaia RGB Thick disk & -0.39 $\pm$ 0.18 & -0.11 $\pm$ 0.26 
    & 26.73 $\pm$ 0.36 & 27.22 $\pm$ 0.50 & 101016 $\pm$ 3103 & 98312 $\pm$ 3991 \\ \hline
    Gaia$\times$RAVE RGB thin disk  & \multicolumn{2}{c}{0.02 $\pm$ 0.55} & 
    \multicolumn{2}{c}{10.74 $\pm$ 0.73} & \multicolumn{2}{c}{7849 $\pm$ 410} \\  
    Gaia$\times$RAVE RGB Thick disk & \multicolumn{2}{c}{-1.13 $\pm$ 0.46} & 
    \multicolumn{2}{c}{28.34 $\pm$ 0.88} & \multicolumn{2}{c}{14251 $\pm$ 905} \\ \hline
    \end{tabular}
    \caption{$V_Z$ mean $\mu$ and dispersion $\sigma$ values from a Gaussian fit (Figure \ref{fig:histovz}) with their respective errors, for the samples studied in this work.}
    \label{tab_gfitvz1}
\end{table}

\begin{figure}[ht!]
\plottwo{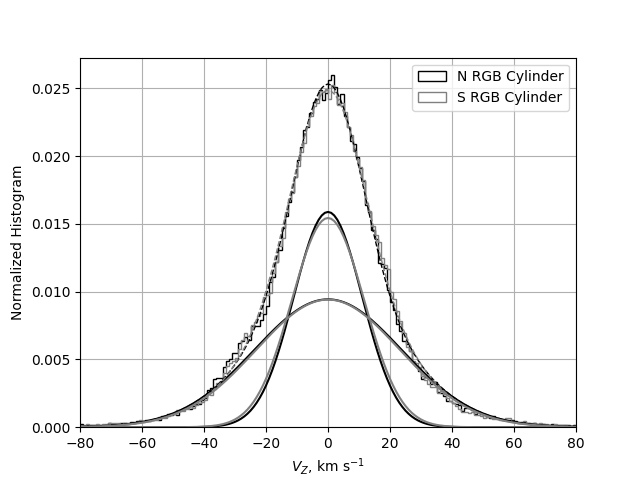}{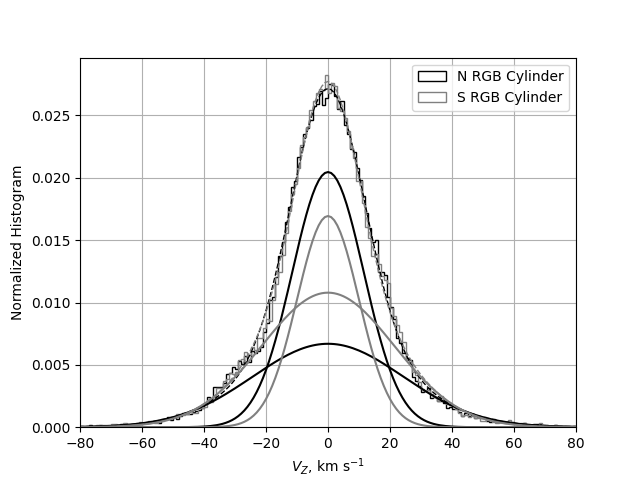}
\caption{Histogram and Sum-of-two-Gaussians fit for $V_Z$, for the samples studied in this work. Left panel is for $|Z|\leq 500$ pc and right one for $|Z|\leq 200$ pc. See also Table \ref{tab_gfitvz2}. \label{fig:fit2g0}}
\end{figure}

A proper estimate of the proportion of Thick and thin disk stars must be performed on the same volume by fitting the $V_Z$ normalized histogram with the sum of two Gaussians, which for simplicity can be assumed both to have mean zero and each its own dispersion, the smaller one of the thin and the larger one of the Thick disk population, including a factor that represents the proportion of Thick disk stars within the whole volume sampled. We did so with the 2 km s$^{-1}$ bins $V_Z$ normalized histogram, with the function {\tt\string optimize.curve\_fit} from the Python {\tt\string scipy} package. Figure \ref{fig:fit2g0} shows the obtained results for the Gaia RGB sample ($|Z|<500$ pc) and for a sub-sample of it up to $|Z|<200$ pc. Table \ref{tab_gfitvz2} shows the results for samples of various heights. 

\begin{table}[ht]
    \centering
    \begin{tabular}{cccccccc}
    Gaia Cylinder RGB & \multicolumn{2}{c}{$\sigma_{thin}$} & \multicolumn{2}{c}{$\sigma_{Thick}$} & \multicolumn{2}{c}{Thick Disk $\%$} \\ \hline    
    Sample height & North & South & North & South & North & South \\ \hline\hline
    $|Z|<500$ pc & 11.09 $\pm$ 0.28 & 11.78 $\pm$ 0.33 
    & 23.90 $\pm$ 0.64 & 23.49 $\pm$ 0.73 & 55 $\pm$ 3 & 53 $\pm$ 4 \\
    $|Z|<200$ pc & 11.68 $\pm$ 0.28 & 10.25 $\pm$ 0.31 
    & 24.76 $\pm$ 1.22 & 21.81 $\pm$ 0.72 & 39 $\pm$ 4 & 55 $\pm$ 4 \\
    $|Z|<100$ pc & 11.51 $\pm$ 0.31 & 10.15 $\pm$ 0.29 
    & 24.27 $\pm$ 1.31 & 21.73 $\pm$ 0.77 & 39 $\pm$ 4 & 51 $\pm$ 4 \\
    $|Z|<\;\;50$ pc  & 10.90 $\pm$ 0.32 & 10.32 $\pm$ 0.31 
    & 23.19 $\pm$ 1.12 & 22.02 $\pm$ 0.89 & 44 $\pm$ 4 & 49 $\pm$ 4
    \end{tabular}
    \caption{$V_Z$ dispersion $\sigma$ values and errors from the sum of two zero-centered Gaussians fit (Figure \ref{fig:fit2g0}).}
    \label{tab_gfitvz2}
\end{table}

\begin{figure}[ht!]
%\plottwo{fig_v2040_bump_500.png}{fig_v2040_bump_200.png}}
%   \caption{Residuals from the $V_Z$ sum-of-two-Gaussian fit.
%   Left panel is for $|Z|\leq 500$ pc and right one for $|Z|\leq 200$ pc.} 
\plottwo{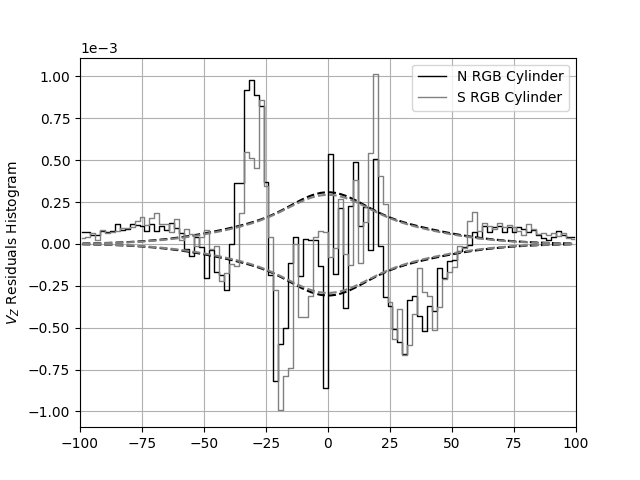}{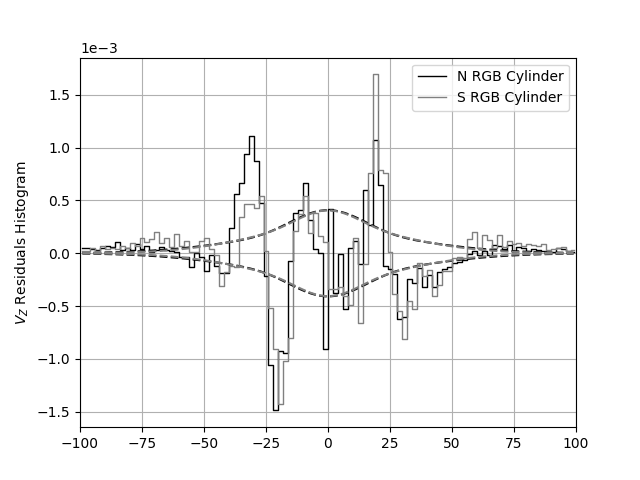}
\caption{Residuals from the $V_Z$ sum-of-two-Gaussian fit for the $|Z|\leq 500$ pc sample (left panel) and the $|Z|\leq 200$ pc sample (right panel). Dashed lines mark the corresponding Poisson noise (1$\sigma$) for each histogram bin. The excess at $-40<V_Z<-20$ and dearth at $25<V_Z<50$ visibly extend beyond the expected noise. Another smaller excess unaccounted for by the fit is also visible at larger velocities in the $|Z|\leq 500$ pc sample.} 
\label{fig:histores}
\end{figure}

A few interesting results emerge: 1) Around half of the cylinder samples are in fact Thick disk stars (see Table \ref{tab_gfitvz2}); 2) There is a North-South number of stars asymmetry more visible in the $|Z|<200$ pc cylinder sample (see Figure \ref{fig:fit2g0} right panel); and 3) There is bit of an excess of stars at $-40<V_Z<-20$ and a dearth at $25<V_Z<50$ more visible when plotting the residuals between the $V_Z$ histogram and the fitted sum of two Gaussians (see Figure \ref{fig:histores}), which occurs similarly in both North and South and becomes more evident in the $|Z|<500$ pc samples. As for point 2), from the sum-of-two-Gaussians fit we estimate that in the $|Z|<200$ pc sample there is a $27\%$ (North) and $42\%$ (South) contamination by Thick disk stars in the thin disk ($|V_Z|<15$ km s$^{-1}$) samples, which levels up to 41\% in both North and South for $|Z|<500$ pc sample (see Appendix \ref{app_thickcontam} for calculations). As for point 3), the residuals in Figure \ref{fig:histores} do not look random with $V_Z$ and the above mentioned excess and dearth of stars go clearly beyond the Poisson noise computed for each histogram bin, by the properly normalized square root of the fit at the bin mid value (statistically, residuals are expected to be within this noise 68\% of the time). An additional but much smaller excess unaccounted for by the fit is also visible at larger velocities in the $|Z|\leq 500$ pc sample, which is probably caused by halo stars contamination.

\begin{figure}[ht!]
\plottwo{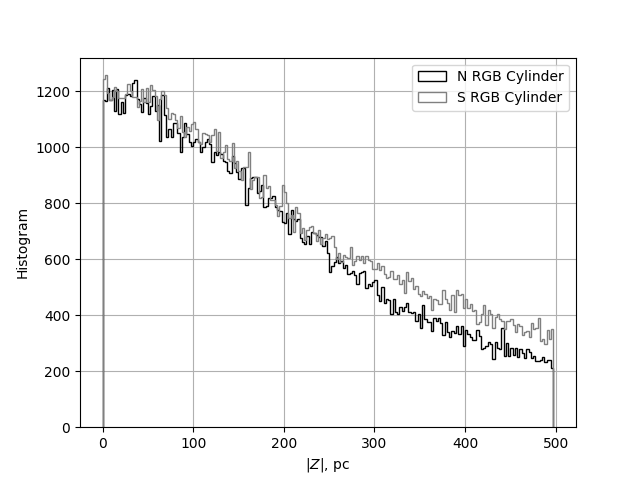}{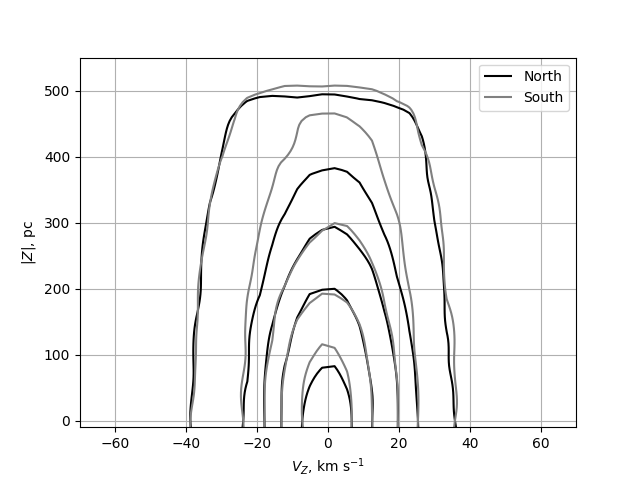}
\caption{Left: Histogram of $|Z|$ for the Gaia Cylinder RGB North and South samples. 
Right: Kernel Density Estimation for $|Z|$ vs. $V_Z$ on the Gaia Cylinder RGB sample. This was implemented using the function \texttt{kdeplot} of Python Seaborn package, isodensity levels shown are 10\% (outermost) to 90\% (innermost) in steps of 20\%.} 
\label{fig_histozz}
\end{figure}

Knowing that the scale height of the Thick disk is significantly larger than the thin disk one ($900 \pm 180$ pc vs. $300 \pm 50$ pc as summarized by \citet{2016ARA&A..54..529B}), it is expected that the proportion of Thick disk stars decreases at smaller heights samples, just as the North samples do, this is why we limited our thin disk samples to $|Z|<200$ pc, in an attempt to reduce Thick disk contamination, while the Thick disk samples go up to $|Z|<500$ pc. The histograms of $|Z|$ for each North and South Gaia Cylinder RGB samples show the South sample number of stars decreases more slowly for $Z\gtrsim 250$ pc (see Figure \ref{fig_histozz}, left panel), yet the excess of Thick disk stars with respect to the thin disk sample in Table \ref{tab_gfitvz2} was detected at heights less than 200 pc. 

A kernel density estimate plot $|Z|$ vs. $V_Z$ (see Figure \ref{fig_histozz}, right panel) of the Gaia RGB sample clarifies these findings. The following two features emerge: 1) the outermost and lowest (10\%) iso-density level for both hemisphere extends farther in {\bf distance} at $-40<V_Z<-20$ than at $25<V_Z<50$. This explains the ``bump'' seen at $-40<V_Z<-20$ and the ``dip'' at $25<V_Z<50$ in Figure \ref{fig:histores}. 2) There are two visible excesses of stars in the South hemisphere that extend farther in height from the Galactic plane than its northern counterpart at iso-density levels 30\% ($300\lesssim |Z|\lesssim 450$) and 90\% ($|Z|\lesssim 100$ pc). The first of these excesses is compatible with Figure \ref{fig_histozz} (left panel) South vs. North excess for $|Z|>250$ pc, while the second one may explain the excess of Thick disk stars in the $|Z|<200$ pc cylinder seen in the right panel of Figure \ref{fig:fit2g0}. 

From Figure \ref{fig:histores} we estimate the number of stars present in the ``bump'' and missing in the ``dip'' at the aforementioned $V_Z$ values, to be as follows: For the 500-pc samples, 1254 and 1319 North stars are in the ``bump'' and missing from the ``dip'', from which at most 265 and 368 stars can be taken as noise, respectively; 768 and 1451 South stars are in the ``bump'' and missing from the ``dip'', from which at most 276 and 379 stars can be taken as noise, respectively. For the 200-pc samples these numbers are respectively for the North: 754 and 576 with a maximum possible noise of 188 and 258 stars, and the South : 338 and 711, with maximum noise of 198 and 268 stars. As seen from the histograms above, these stars are just a small percentage of the whole RGB sample, amounting to only 0.5 to 1\% of the stars, yet their presence is visible in all the plots analyzed.

\begin{figure}[ht!]
\plottwo{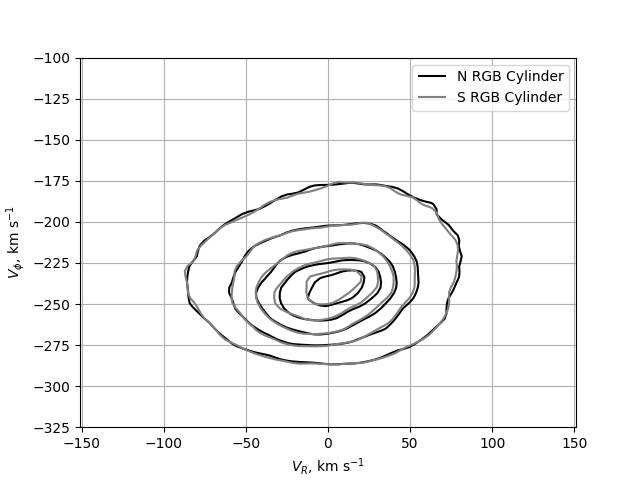}{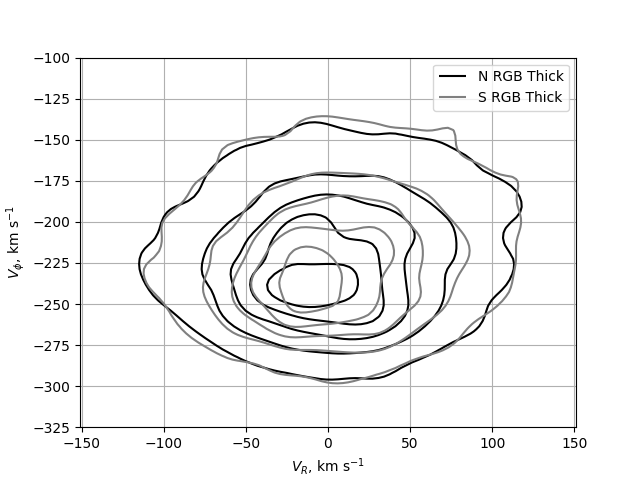}
\caption{Left: Kernel Density Estimation for $V_\phi$ vs. $V_R$ on the Gaia RGB North and South samples. Parameters as in the previous figure. Right: Same for the Gaia Thick disk samples.}
\label{fig_kde}
\end{figure}

In light of the $V_Z$ results, we evaluated also the $V_R$ and $V_\phi$ velocities to look for such features in those velocities. When analyzing the whole RGB cylinder, with both thin and Thick disk populations mixed, it is hard to see an asymmetry except possibly at the innermost density level (see Figure \ref{fig_kde}, left panel). But the same plot for the Gaia Thick disk sample, shows some localized asymmetries (see Figure \ref{fig_kde}, right panel). We do not attempt this analysis in the Gaia thin disk sample as we have proven already that about half of that sample is composed of Thick disk stars. 

\section{Discussion}

Using RAVE data, \citet{williams2013wobbly} studied the kinematical properties of red clump giants within a few kpc radius and height volume around the Sun. They found differences between the North and South in the radial velocity streaming motions, $V_R$. \citet{williams2013wobbly} found also a surprising complex behavior of $V_Z$ velocity component in solar neighborhood. Interior to the solar circle, stars move upwards above the plane and downwards below the galactic plane. Exterior to the solar circle the stars both above and below the plane move towards the galactic plane with velocities up to $|V_Z|=17$ km s$^{-1}$. The authors interpret such behavior of $V_Z$ as a wave of compression and rarefaction in the direction perpendicular to the Galactic plane. Such wave, as \citet{williams2013wobbly} suggest, could be caused either by a recently engulfed satellite, or by the disk spiral arms.

Our analysis also shows in the solar neighborhood an anomaly in $V_Z$-velocity field. The distribution of $V_Z$ velocity of red giant stars has an excess of stars with velocities -40 to -20 km s$^{-1}$ (see Figures \ref{fig:fit2g0} and \ref{fig_histozz}-right panel) and a dearth of stars with $V_Z$ velocities 25 to 50 km s$^{-1}$. Such features are also seen  in Figure \ref{fig_kde} that shows $Z$ vs $V_Z$ iso-density contours for Thick disk stars both in the southern and northern galactic hemispheres. At the same time  $V_R$ and $V_{\phi}$ velocity components of red giant stars do not have such an anomaly. 
Several mechanisms can be responsible for the observed peculiarity in the $V_Z$ stellar velocity distribution in the solar neighborhood. The disk is strongly affected by the Galactic bar and spiral structure \citep{2018Natur.561..360A}. Non-equilibrium phase mixing can occur due to the tidal disturbance of the Galactic disk by the crossing of a Sagittarius-like satellite with mass $\sim 3 \times 10^{10}$ M$_\odot$ \citep{2019MNRAS.486.1167B}. Further study is needed to identify mechanism responsible for these features.

\citet{lee2011formation} used a sample of 17,277 G-dwarfs with measured $[\alpha/Fe]$ ratio and chemically separated the disk onto thin and Thick disk populations. \citet{lee2011formation} used also proper motion information of their sample with typical proper motion errors about 3-4 mas/yr together with distances to individual stars estimated with help of calibrated stellar isochrones. Using these data, \citet{lee2011formation} measured the velocity lag between chemically separated thin and thick dicks to be nearly constant $\sim$30 km s$^{-1}$ at any given distance $|Z|$ from the galactic plane. Our estimate of the velocity lag between the thin and Thick disks is about 14 km s$^{-1}$. We checked an influence of metallicity information on the value of the velocity lag. Cross-matching the sample with the RAVE DR5 catalog and taking into account metallicity information shows that metallicity does not change essentially the value of the velocity lag of the Thick disk in the solar neighborhood.

Recently, \citet{anguiano2020stellar} used stellar metallicity information to discriminate the three primary stellar populations: thin disk, Thick disk and halo.  Chemistry-based selection of the stars belonging to different galactic subsystems allowed \citet{anguiano2020stellar} to select 211,820 stars associated with the Milky Way thin disk, 52,709 stars associated by their abundances to the Thick disk, and 5,795 stars belonging to the halo population. The sample of stars used by \citet{anguiano2020stellar} spans approximately $6<R<10$ kpc in Galactocentric cylindrical radius and $-1<Z<2$ kpc in $Z$-coordinate. \citet{anguiano2020stellar} found that chemically selected thin disk has velocity dispersion $(\sigma_R, \sigma_\phi, \sigma_Z)$ of $(36.81, 24.35, 18.03) \pm (0.07, 0.04, 0.03)$ km s$^{-1}$. For the thick disk, the authors get a velocity dispersion $(\sigma_R, \sigma_\phi, \sigma_Z)$ of $(62.44, 44.95,41.45) \pm (0.21, 0.15, 0.15)$ km s$^{-1}$. The mean rotational velocity of the chemically selected Thick disk according to \citet{anguiano2020stellar} is equal to $191.82 \pm 0.24$ km s$^{-1}$ and their value of asymmetric drift between Thick and thin disks is about 30 km s$^{-1}$. Our analysis of the kinematical properties of red giant stars selected  close to the Sun gives a smaller value of asymmetric drift about 19 km s$^{-1}$. Discrepancy between the values of the velocity lag of the Thick disk may be due to the fact that the sample of stars selected by \citet{anguiano2020stellar} has a much larger volume compared to our sample selected in a close proximity to the Sun. Also, \citet{anguiano2020stellar} did not discuss the issue of completeness of their sample which can be essential in determination the relative velocity lag between the subsystems.
 
 A more serious discrepancy between our study and the results of \citet{anguiano2020stellar} appears in the estimate of the proportion of stars that belong to the different subsystems of the Galaxy. \citet{anguiano2020stellar} estimate that in their data set 81.9 percent of stars belongs to the thin disk, 16.6 percent are the Thick disk stars, and about 1.5 percent of the stars belong to the Milky Way halo. The local Thick-to-thin density normalization $\rho_{Thick}$ / $\rho_{thin}$ was estimated by \citet{anguiano2020stellar} to be about 2 percent.  We find that in our complete kinematically selected sample of stars the ratio of local Thick-to-thin number density is about 90 percent (from the last row in Table \ref{tab_gfitvz2}). Two comments should be made here. First, as \citet{kawata2016milky} have noticed, the Thick disks that are selected chemically or kinematically are strictly speaking different objects. Our criterion of selection allows to choose the complete sample of stars that have in the solar neighborhood the Thick disk {\it kinematics}, i.e., these stars will deviate in the direction perpendicular to the Galactic plane by 1-2 kpc in a near future. The stars selected this way represent the wings of the Thick disk vertical velocity distribution. This result was confirmed independently by fitting the complete sample of stars with the sum of two Gaussian distribution. By incorporating metallicity information for our sample and dividing the sample of stars onto the thin ($[Fe/H]>-0.4$) and the Thick ($-1<[Fe/H]<-0.4$) disk stars does not change the kinematical properties of the thin and Thick disks in solar neighborhood.
 
 \citet{everall1,everall2} used Gaia photometry and astrometry to estimate the spatial distribution of the Milky Way disk at the Solar radius. To correct  for  sample incompleteness, 
 they used a solution for selection function for GAIA source catalog  \citet{evebou,boueve} and  recovered the densities of the Thin and of the Thick disks in Solar neighborhood, and the scale heights of the vertical density distributions in both disks. \citet{everall1} find a ratio of the Thick-to-Thin local density of 0.147 $\pm$ 0.005, and a value of surface density of the stellar disk of 23.17 $\pm$ 0.08 (stat) $\pm$ 2.43 (sys) M$_{\odot}$ pc$^{-2}$. Their first value is considerably lower than our estimate of the local ratio of Thin-to-Thick densities obtained from a complete sample of red clump stars, and their second value is also lower than previous estimates of the surface density of stars in Solar neighborhood, as concluded by them.

 Our finding that the Milky Way thick disk has mass comparable to that of the thin disk concurs with the result of \citet{kawata2016milky}. These authors determined the Milky Way star formation history using the imprint left on chemical abundances of long-lived stars. \citet{kawata2016milky} find that the  formation of the Galactic Thick disk occurred during an intense star formation phase between 9.0 and 12.5 Gyr ago that was followed by a dip in star formation rate lasting about 1 Gyr. The intense phase of star formation in the past of the Milky Way galaxy resulted in the formation of a massive Thick galactic disk. In a another paper, \citet{lehnert2014milky} compared the star formation history of the Milky Way galaxy with the properties of distant disk galaxies. They found that during the first 4 Gyr of its evolution, the Milky Way formed stars with a high rate ($\sim 0.6 M_{\odot}$ yr$^{-1}$ kpc$^{-2}$) resulting in the formation of the thick Milky Way disk with a mass approximately equal to that of the thin Milky Way disk.

An additional piece of information comes from the comparison of the Milky Way Thick disk stellar eccentricity distribution with that presented for simulated disks formed via accretion, radial migration, and gas-rich mergers \citep{wilson2011testing}. The authors find that the broad peak at moderately high eccentricities in the accretion model is inconsistent with the relatively narrow peak at low eccentricity observed in the Milky Way Thick disk, which indicates, that the Galactic thick disk was formed predominantly {\it in situ}.

The above mentioned results are in agreement with recent high-resolution NEWHORIZON and GALACTICA simulations \citep{park2021exploring}. These authors find that being spatially separated, the two disks contain overlapping components, so that even in the Galactic mid-plane Thick disk stars contribute on average $\sim$ 30 percent in total density of stars, and about 10 percent in their luminosity. \citet{park2021exploring} conclude that spatially defined thin and Thick disks are not entirely distinct components in terms of formation process. The two disks represent parts of a single disk that evolves with time due to continuous star formation and disk heating, which in fact finds confirmation in our analysis of kinematics of stars in solar neighborhood. 

\section{Conclusions}

Using a complete sample of 296,879 RGB stars distributed in a cylinder centered at the Sun with a 1 kpc radius and half-height of 0.5 kpc, we study the kinematical properties of the Milky Way disk in the solar neighborhood. Analysis of kinematical properties of RGB stars in solar neighborhood was done with help of a two-component fit to the velocity distribution of $V_Z$ velocity component. Our results can be summarized as follows.

1.	The kinematical properties of the selected stars point at the existence of two distinct components: the thin disk with mean velocities $V_R$, $V_{\phi}$, $V_Z$ of -1, -239, 0 km s$^{-1}$, and velocity dispersions $\sigma_R$, $\sigma_{\phi}$, $\sigma_Z$ of 31, 20 and 11 km s$^{-1}$, correspondingly. The Thick disk component has, on the other hand, mean velocities $V_R$, $V_{\phi}$,  $V_Z$ of +1, -225, 0 km s$^{-1}$, and  velocity dispersions $\sigma_R$, $\sigma_{\phi}$,  $\sigma_Z$ of 49, 35, and 22 km s$^{-1}$. Completeness of our RGB sample of stars allows to estimate the density ratio of the thin and Thick disks in the solar neighborhood.  We find that Thick disk stars comprise about half the stars of the disk. Such high density of the stars with Thick disk kinematics points at an 
{\it in situ} rather than {\it ex situ} formation of the Thick disk.

2. 	$V_Z$ velocity field has {\bf small but real} anomaly in the solar neighborhood. Velocity distribution of red-giant stars in the direction perpendicular to the galactic plane has an excess of stars with $V_Z$ velocities of -40 to -20 km s$^{-1}$ and a dearth of stars with $25<V_Z<50$ km s$^{-1}$. Anomaly is observed both in the Northern and in the Southern Galactic hemispheres.
%and points at the presence of ***a stream***, running in %nearly perpendicular to the Galactic disk.

\begin{acknowledgments}

VK acknowledges support from Russian Science Foundation (Project No- 18-12-00213-P).
%This work made use of the programming language Python (https://www.python.org/).
\end{acknowledgments}


\vspace{5mm}

\software{Python v. 3.8.10 (\url{https://www.python.org/}), 
          Topcat \citep{2005ASPC..347...29T}, SciPy \citep{2020SciPy-NMeth},
          Seaborn \citep{Waskom2021}}



\appendix

\section{Appendix A - GAIA EDR3 and RAVE DR5 crossmatch}\label{app_xmatch}

The main reported results of this investigation are based exclusively on kinematical data from GAIA EDR3, yet a small subsample was crossmatched with RAVE DR5, in order to have RAVE's abundance $[Fe/H]$ as an additional chemical constrain, and use these more refined samples to check our main results. The 5th installment of the RAVE catalogs has 520,701 entries but only 457,588 unique stars were observed \citep{2017AJ....153...75K}. A non-negligible portion of those entries are several observations of the same star, and these entries may or not have the same RAVEID in the catalog, as the ID is based on coordinates, and these can change slightly from one observation to other. We first reject all entries that have a repeated RAVEID, in some cases they showed significant variations in their RAVE radial velocity. We then deleted all clusters or groupings of stars that were within 1 arcsecond from each other. This final clean version of RAVE DR5 contains 405,231 stars. 

After some tests, a threshold of 1.7 arcseconds for positional matching by angular distance between the clean version of RAVE and Gaia reported coordinates was set. Similar to \citet{2018RNAAS...2..194S}, we found a subset of stars (536 of them) with a radial velocity offset $\Delta RV$ of about +100 km s$^{-1}$ with respect to GAIA. \citet{2018RNAAS...2..194S} indicates these stars were located close to the plate edges when observed but since the catalog does not contain any information about this, there is no way to discard them safely. 
At this point, for each Gaia star, we take all RAVE DR5 matches within 1.7 arcseconds and for which  $\Delta RV<90$ km s$^{-1}$. A few cases of multiple matches were discarded (96 Gaia stars were matched to two different RAVE stars, 409 RAVE stars were matched to two different Gaia sources). The final match catalog RAVE x GAIA has 356,270 sources (88\% of the clean version of the RAVE catalog). An additional piece of information indicating these matches are good, is the comparison between GAIA $G$ magnitude (broad band filter 400-850 nm, centered at 673 nm) and APASSDR9 $R$ magnitude (600-750 nm filter). About $\sim 18,000$ RAVE stars do not have $R$ magnitude, but for the rest of the sample ( $\sim 338,000$ stars), $G-R$ values are well centered around cero, with some dispersion (differences within 1 magnitude for the faintest stars). 

Some low signal-to-noise ratio RAVE entries (STN\_SPARV $\sim $ 1) exhibit way too large uniformly distributed  $\Delta RV$ within the limit imposed, a typical telltale of mismatches. Later on when selecting the samples for this research, these stars did not make it afterwards. In any case, the poor radial velocity agreement for the low signal-to-noise data may perfectly be caused for many of them by just bad quality RAVE data. 

\section{Appendix B - Incompleteness estimation} \label{app_complete}

When a sample $\{X_i\}_{i=1}^n$ is distributed uniformly within a given interval $[0,1]$, then
its empirical CDF is very close (within statistical noise) to the identity function in that interval. 
In other words, $p_i\sim X_i$ for $i=1,\dots, n$, where $p_i$ is the corresponding percentile. 
A systematic difference between both will imply incompleteness or a not homogeneous sampling of the underlying distribution. As Figure \ref{fig:cdf} suggest for the Gaia samples, the percentile of the data is generally below the data, in other words some stars {\bf should occupy higher percentiles than they currently do for a uniform distribution}. In general, we can compute the $\hat{X}\neq 1$ such that the area of the data CDF would correspond to the uniform distribution CDF in the interval $[0,\hat{X}]$, in other words, the value up to which the current $n$ data points would homogeneously sample a uniform distribution, then
\begin{equation}
\sum_{i=1}^{n-1} (X_{i+1}-X_i)p_i=\frac{\hat{X}^2}{2} \Longrightarrow \hat{X}=\sqrt{2\sum_{i=1}^{n-1} (X_{i+1}-X_i)p_i}
\end{equation}
If we sample the same population but over the interval $[0,1]$ with $N\neq n$ points, then trivially $\frac{N}{1}=\frac{n}{\hat{X}}$. Finally, the incompleteness percentage can be computed as
\begin{equation}
\mbox{Incompleness \%}=\left(1-\frac{n}{N}\right)\times 100= (1-\hat{X})\times 100=
\left(1-\sqrt{2\sum_{i=1}^{n-1} (X_{i+1}-X_i)p_i}\right)\times 100    
\end{equation}

\section{Appendix C - Thick disk contamination in thin disk sample} \label{app_thickcontam}
For a normal distribution $N(\mu,\sigma)$, the cumulative distribution function CDF is given by
\begin{equation}
    CDF(x)=\frac{1}{2}\left[1+\mbox{erf}\left(\frac{x-\mu}{\sigma\sqrt{2}}\right)\right]
\end{equation}
The area under $N(0,\sigma)$ in the interval $[-15,15]$ is given by
\begin{equation}
    CDF(15)-CDF(-15)=\frac{1}{2}\left[1+\mbox{erf}\left(\frac{15}{\sigma\sqrt{2}}\right)\right]-
    \frac{1}{2}\left[1+\mbox{erf}\left(\frac{-15}{\sigma\sqrt{2}}\right)\right]=
    \mbox{erf}\left(\frac{15}{\sigma\sqrt{2}}\right)
\end{equation}
Let be $0\leq p_T\leq 1$ the proportion of Thick disk stars whose velocity distribution is $N(0,\sigma_T)$, that is mixed with a population of Thin disk with velocity distribution $N(0,\sigma_t)$. When restricted to the interval $[-15,15]$, which we used to define the Thin disk sample, the contamination by Thick disk stars is given by
\begin{equation}
    \frac{p_T\mbox{erf}\left(\frac{15}{\sigma_T\sqrt{2}}\right)}
    {p_T\mbox{erf}\left(\frac{15}{\sigma_T\sqrt{2}}\right)+
    (1-p_T)\mbox{erf}\left(\frac{15}{\sigma_t\sqrt{2}}\right)}
\end{equation}


\bibliography{kvieira}{}
\bibliographystyle{aasjournal}



\end{document}